\documentclass[preprint,showpacs,preprintnumbers,amsmath,amssymb]{revtex4}
\usepackage{ifpdf}
\ifpdf
  \usepackage[pdftex]{graphicx}
  \DeclareGraphicsExtensions{.jpg, .png , .pdf}
\else
  \usepackage[dvips]{graphicx}
  \DeclareGraphicsExtensions{.eps, .ps, .eps.gz, .ps.gz}
\fi

\usepackage{dcolumn}% Align table columns on decimal point
\usepackage{bm}% bold math

\begin{document}

%\preprint{APS/123-QED}

\title{The domain wall spin torque-meter\\}% Force line breaks with \\

\author{I.M. Miron}
\author{P.-J. Zermatten}
\author{G. Gaudin}
\author{S. Auffret}
\author{B. Rodmacq}
\author{A. Schuhl}

\affiliation{SPINTEC, CEA/CNRS/UJF/GINP, INAC, 38054 Grenoble Cedex 9, France  %\\
}
\date{\today}% It is always \today, today,

\begin{abstract}

We report the direct measurement of the non-adiabatic component of the spin-torque in domain walls. Our method is independent of both the pinning of the domain wall in the wire as well as of the Gilbert damping parameter. We demonstrate that the ratio between the non-adiabatic and the adiabatic components can be as high as 1, and explain this high value by the importance of the spin-flip rate to the non-adiabatic torque. Besides their fundamental significance these results open the way for applications by demonstrating a significant increase of the spin torque efficiency. 

\end{abstract}

\pacs{72.25.Rb,75.60.Ch,75.70.Ak,85.75.-d}

\maketitle
The possibility of manipulating a magnetic domain wall via spin torque effects when passing an electrical current through it opens the way for conceptually new devices such as domain wall shift register memories\cite{parkin}. Early spin-torque theories\cite{berger84,thiaville-adiabatic,Slonczewski96} were based on a so called adiabatic approximation which assumed that the incoming electron's spin follows exactly the magnetization as it changes direction within the domain wall. Nevertheless, the observed critical currents needed to trigger the domain wall motion were lower than the value predicted within this framework\cite{klaui1}. As first predicted by Zhang\cite{zhang}, the existence of a non-adiabatic term in the extended Landau-Lifshitz-Gilbert equation leads to the vanishing of the intrinsic critical current. The action of this non-adiabatic torque on a DW is expected to be identical to that of an easy axis magnetic field. Micromagnetic simulations have been used to predict the velocity dependence on current for a DW submitted to the action of the two components of the spin-torque\cite{thiaville-nonadiabatic}.
The quantitative measurement of this non-adiabatic torque can be achieved either by demonstrating the equivalence of field and current in a static regime, or by observing the complex dynamic behavior\cite{thiaville-nonadiabatic}. The main difficulty of these measurements comes from the pinning of the DW by material imperfections. It masks the existence of the intrinsic critical current, and in addition, above the depinning current, obscures the DW velocity dependence on current. Moreover, most of the DW velocity measurements were done using materials with in-plane magnetization\cite{klaui1,beach-current,hayashi,jubert,vernier}, where the velocity can also depend on the micromagnetic structure of the wall\cite{nakatani} (transverse wall or vortex wall). Despite the simpler micromagnetic structure of the DWs, very few results were reported\cite{ravelosona} for Perpendicular Magnetic Anisotropy (PMA) materials. In this case the intrinsic pinning is much stronger, probably due to a local variation of the perpendicular anisotropy.
Up to now, none of the measurements were able to clearly evidence the equivalence between field and current, nor to reproduce the predicted dynamic behavior; hence the value of the non-adiabatic torque is still under debate.

In this letter we use a novel approach for the measurement of the non-adiabatic component of spin-torque. Instead of measuring the DW velocity, we perform a quasistatic measurement of its displacement under current and magnetic field. 
In principle this method is similar to any quasi-static force measurement: a small displacement is created, first with the unknown force and then with a known reference force. In our case the unknown force is caused by the electric current passing through the DW while the reference force is due to an applied magnetic field. By comparing the two displacements one directly compares the applied forces. Due to the high sensitivity of our method (able to detect DW motion down to $\sim$10$^{-2}$nm\cite{sup}) we can study the displacement of the DW inside its pinning center. Since the measurement relies on the comparison to a reference force, the method is independent of the strength of the pinning. Moreover, as the field and current are applied quasi-statically, the damping parameter does not play any role.

\begin{figure}
%\scalebox{0.4}{
\includegraphics[angle=0,width=1\linewidth]{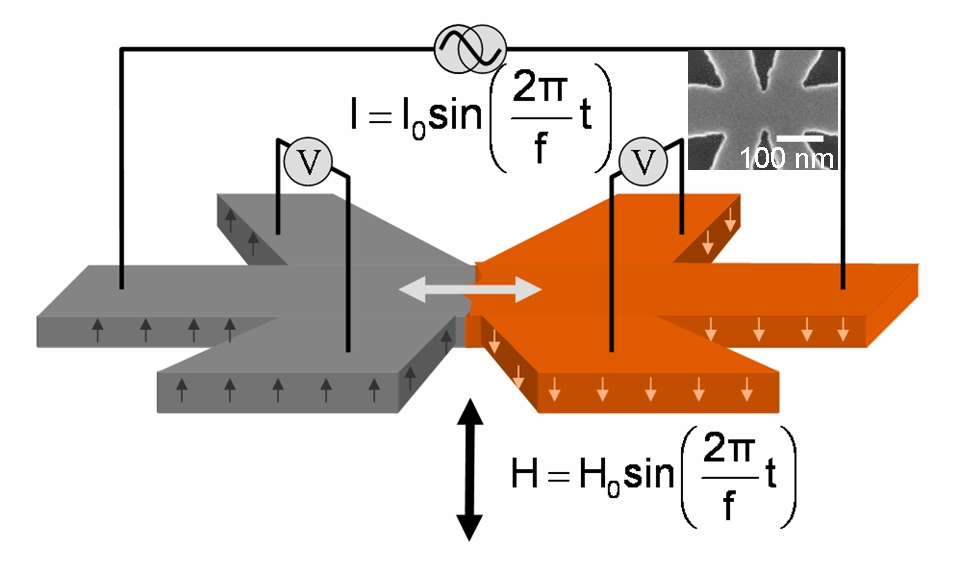}
\caption{Schematic representation of the experimental setup. The inset shows an SEM picture of a sample.}
\label{1}
\end{figure}

According to recent theories\cite{zhang,vanke+viret,piechon} that derived the value of the spin torque, $\beta$ (the ratio between the non-adibatic and adiabatic torques) is given by the ratio between the rate of the spin-flip of the conduction electrons and that of the s-d exchange interaction. Generally, two conditions must be fulfilled to obtain a high spin-flip rate. First it is necessary to have a strong crystalline field inside the material. The electric fields will yield a magnetic field in the rest frame of the moving electrons. Second, a breaking of the inversion symmetry is needed. Otherwise the total torque of the magnetic field on the electron spin averages out, and the spin-flip may only occur during momentum scattering\cite{fabian-sf}.

In order to highlight these effects we have patterned samples from a Pt\scriptsize3nm\normalsize/ Co\scriptsize0.6nm\normalsize/(AlO\scriptsize x\normalsize)\scriptsize2nm\normalsize  \:layer\cite{manchon}. In this case the symmetry is broken by the presence of the AlO\scriptsize x\normalsize \:on one side of the Co layer, and of the heavy Pt atoms on the other\cite{krupin,cercellier-rash}. We will emphasize the importance of the spin-flip interaction to spin torque by comparing results from these samples with those for samples fabricated from a symmetric Pt\scriptsize3nm\normalsize/ Co\scriptsize0.6nm\normalsize /Pt\scriptsize 3nm\normalsize  \:layer\cite{metaxas}, where a much smaller spin-flip rate is expected. As the only difference between the two structures is the upper layer, we expect similar growth properties for the Co layer. Both samples exhibit PMA and a strong Anomalous Hall Effect (AHE)\cite{canedy-ehe}. 
The films are patterned into the shape depicted in Figure 1. This shape is well suited for a quasi static measurement as a constriction is created by the presence of the four wires used for the AHE measurement (figure 1 inset). This way a DW can be pinned in a position where changes in the out of plane component of the magnetization (i.e. DW motion) can be detected by electrical measurements. A current is passed through the central wire. This current will serve to push the domain wall as well as to probe the eventual displacement. In the case where the DW does not move under the action of the current, the transverse resistance remains unchanged and the voltage measured across the side wires (AHE) will be linear with the current. If the DW moves due to the electric current, the exciting force will create resistance variations, causing a nonlinear relationship between the measured voltage and the applied current.
A simple way to detect such nonlinearities is to apply a perfectly harmonic low frequency (10 Hz) ac current, and look at the first harmonic in the Fast Fourier Transform (FFT) of the measured voltage. Its value is a measure of the amplitude of the DW displacement at the frequency of the applied current.
To quantitatively compare the action of a magnetic field to that of an electric current, the magnetic field is applied at the same frequency and in phase (or opposition of phase) with the electric current. By applying current and field simultaneously, we ensure that their corresponding torques act on the same DW configuration. In addition to the displacement provoked by the current, the field induced displacement will add to the value of the first harmonic, which can be either increased if the field and current push the wall in the same direction, or decreased if they act in opposite directions. 

\begin{figure}
%\scalebox{0.4}{
\includegraphics[angle=0,width=0.8\linewidth]{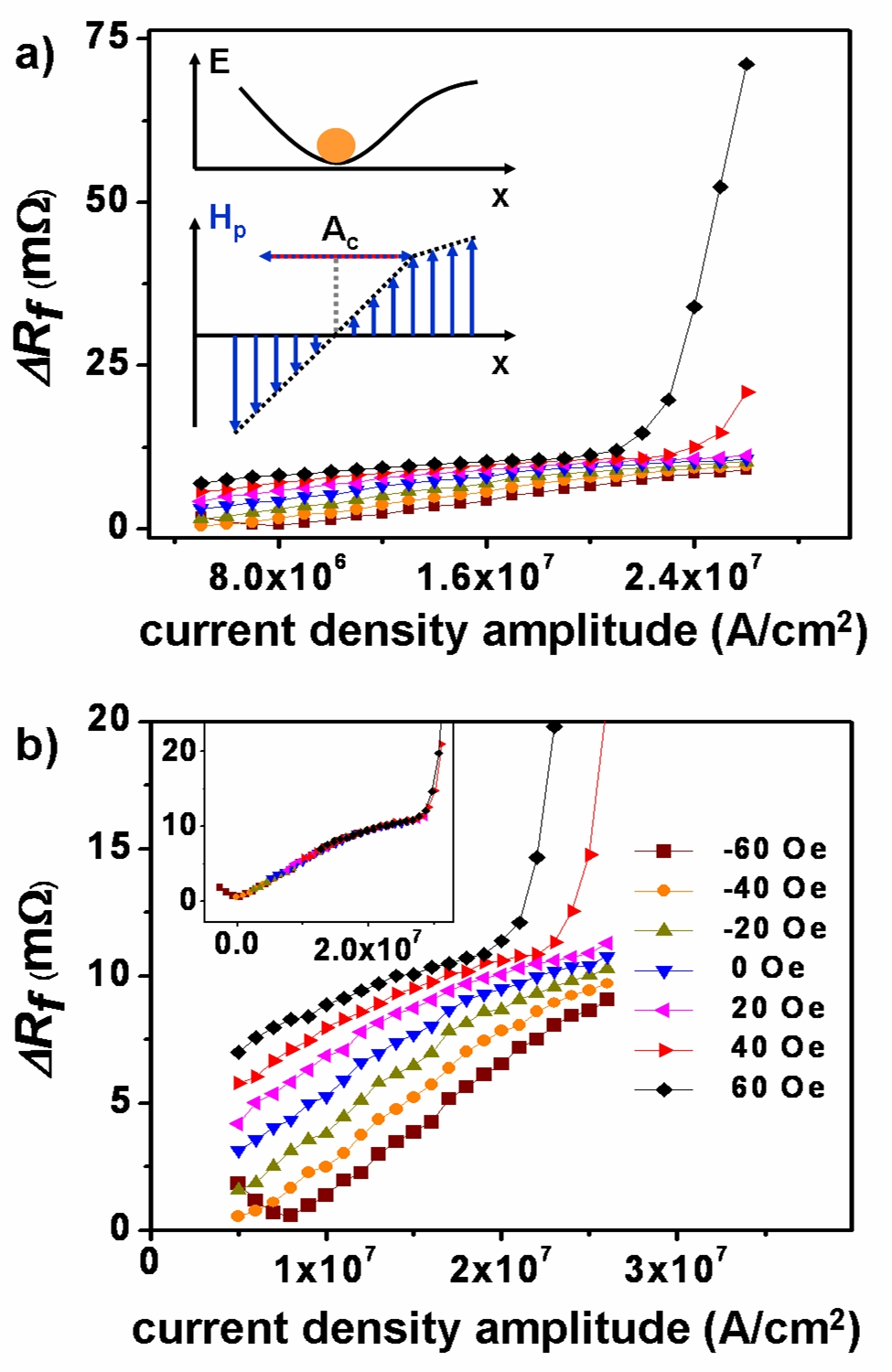}
\caption{(a) Dependence of the resistance variation on the current amplitude for several field amplitudes (Pt/Co/AlO\scriptsize x\normalsize\; sample). The inset shows a possible nonlinear and asymmetric potential well. The energy landscape can be modeled by an effective out of plane magnetic field that has negative values on one side of the equilibrium position and positive values on the other. (b) A zoom on the small amplitude regime. The inset shows the perfect superposition obtained by shifting the curves horizontaly with 1.25 $\cdot$ $10^{5}$Acm$^{-2}$ Oe$^{-1}$.}
\label{1}
\end{figure}
Figure 2 shows the dependence of the resistance variation at the frequency of the current ($\Delta$R\scriptsize f\normalsize) on the current amplitude for different values of the field amplitude. First, at low current and field amplitudes the displacement is almost linear ($\sim$10$^{7}$A/cm$^{2}$), but for higher values, the $\Delta$R\scriptsize f\normalsize \:varies more rapidly.	
A simple estimation based on the value of the resistance variation compared to the total Hall resistance of a cross (1 $\Omega$) yields $\sim$ 1 nm for the maximum amplitude of the DW motion in the first regime and $\sim$ 7 nm for the second regime. 
This behavior can be explained by the anatomy of the local pinning. The local potential well trapping the DW can be considered as a superposition of the geometric pinning \cite{bubble} and intrinsic pinning caused by defects randomly distributed inside the material\cite{ravelosona,metaxas}. Because the potential well for the small scale displacements (below 10nm) is dominated by the random intrinsic pinning rather than geometric pinning (the increase of the length of the DW is small $\sim$ 1\%) in the general case it should be asymmetric. 
We have verified the supposed asymmetry of the effective potential well by applying alongside the ac current and field, a dc bias field that changes the local potential well (inset of figure 3). By varying this field we observed a reduction of the current amplitude needed to access this strongly non-linear regime (figure 3). When the magnetic bias field was reversed this second regime was no longer attained with the available current densities (not shown).
\begin{figure}
%\scalebox{0.4}{
\includegraphics[angle=0,width=0.7\linewidth]{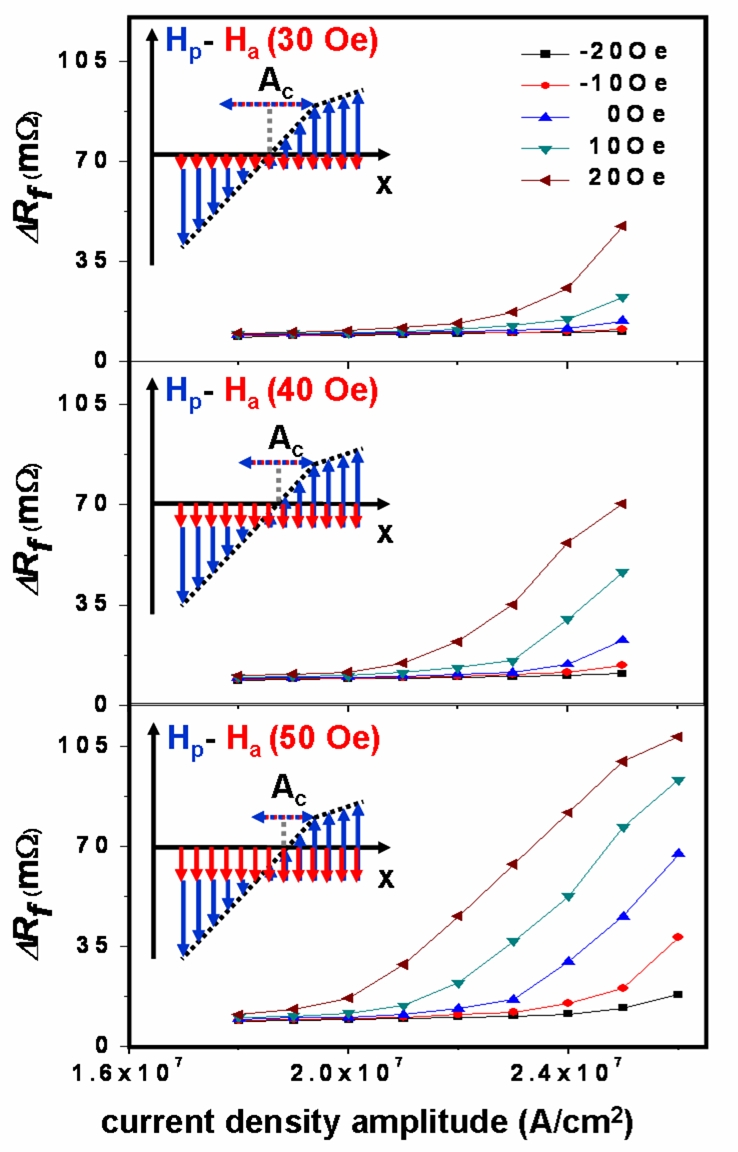}
\caption{The nonlinear regime (Pt/Co/AlO\scriptsize x\normalsize\; sample). When an external bias field is added, the effective pinning field changes (inset) and the nonlinear regime is reached for different current and field amplitudes. However, this does not cause any change in the field to current correspondence: the horizontal distance between the curves remains the same. }
\label{1}
\end{figure}	

The observed dependence of $\Delta$R\scriptsize f\normalsize \;on current and field (figures 2 and 3) is in perfect agreement with the characteristic features of the non-adiabatic component of the spin-torque. First, we do not observe any critical current down to the lowest current value (10$^{6}$A/cm$^{2}$- figure 4 in \cite{sup}). Futhermore, by extrapolating the amplitude of the DW displacement (figure 2), when the current is reduced, the displacement goes to zero as the current goes to zero, in agreement with the absence of the critical current. 

However, the most important feature of the $\Delta$R\scriptsize f\normalsize \; behavior is that the curves obtained for any field amplitude can be obtained from the curve corresponding to zero field just by shifting it horizontally (in current): towards the lower current values when the field and current act in the same direction on the DW and towards higher values when their actions are opposed.
This means that any displacement of the DW can also be achieved with a different current if a magnetic field is added. The difference in current is compensated by the magnetic field. The value of this horizontal shift gives the field to current correspondence. The inset of figure 1b shows that all the curves corresponding to different field amplitudes have the same shape; by shifting them horizontally (using the field-current correspondence), they all collapse on the zero field amplitude curve.
This shows that independently of the direction or strength of the applied current and field, as predicted by the theories, their effect on the DW is fundamentally similar. Moreover, further evidence that this correspondence is intrinsic and not influenced by pinning is that its value remains the same within the different amplitude regimes as well as when the local potential well is tuned by a constant bias field.

Since the motion of the DW is quasi-static the magnetization can be considered to be at equilibrium during motion. In this case the sum of all torques must be zero. 
In order for the DW to remain at rest, the torque from the applied current must be compensated by the torque generated by the magnetic field.
The upturn observed on the -60 Oe curve (figure 2b) determines the position of the zero amplitude point. Note that the position of this point is in perfect agreement with the field to current correspondence obtained from the horizontal shifting of the curves.
By taking into account the micromagnetic structure of the DW (very thin ~5nm Bloch wall) the two torques are integrated over the width of the wall, and by comparing their values (the field torque is easily calculated; \cite{sup}) the non-adiabatic term of the spin-torque is determined. In the case of Pt/Co/AlO\scriptsize x\normalsize \: stacks the current-field correspondence is approximately 1.25 10$^{5}$ A/cm$^{2}$ to 1Oe, corresponding to a value of $\beta$ = 1.

Similar measurements (figure 1 in \cite{sup}) were also performed in the saturated state (without the DW). They confirm that there is no contribution to the signal from the ordinary Hall effect, but indicate a small contribution from thermoelectric effects - the Nernst-Ettingshausen Effect(NEE)\cite{thaler-NEE}.
The contribution from DW motion to $\Delta$R\scriptsize f\normalsize \:is much higher than the NEE for the Pt/Co/AlO\scriptsize x\normalsize \: stack. In the case of Pt/Co/Pt layers we find that the amplitude of the current induced DW motion is much smaller and entirely masked by the NEE. When a DW is moving inside the perfectly harmonic region at the bottom of the potential well, its displacement depends linearly on the applied force. In such a scenario, the current induced DW motion and the NEE are indistinguishable. They both lead to a linear dependence of the $\Delta$R\scriptsize f\normalsize \:response on current. 
The only possibility to separate these effects, for the Pt/Co/Pt layer, is to attain the high amplitude nonlinear regime of DW motion. This is done by keeping the current amplitude constant and varying the field amplitude. When the current and field push the wall in the same direction, the nonlinear regime should be reached for smaller field amplitudes, than if their actions were opposed.
\begin{figure}
%\scalebox{0.4}{
\includegraphics[angle=0,width=0.7\linewidth]{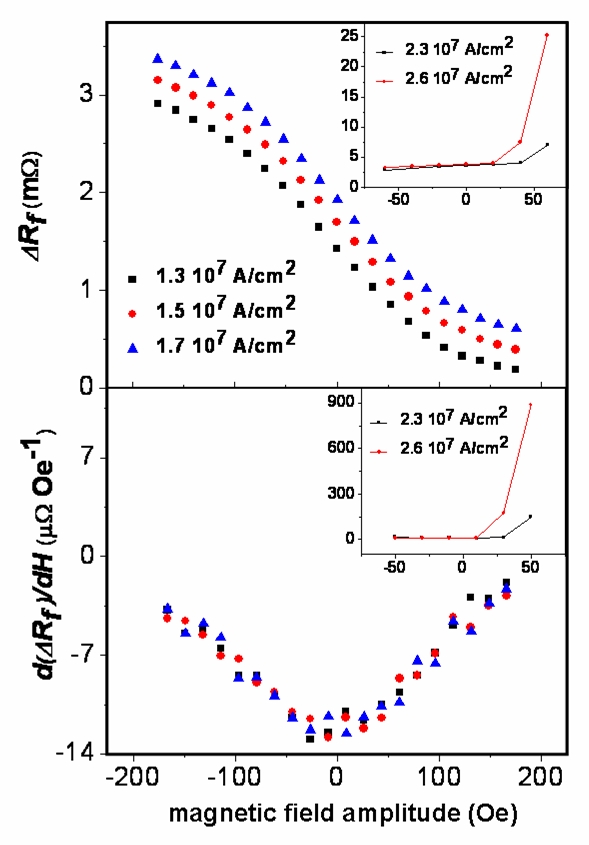}
\caption{The nonlinear response of a DW to magnetic field. (a) $\Delta$R\scriptsize f\normalsize \: vs. the amplitude of the field for three different current densities in the case of Pt/Co/Pt layers (inset Pt/Co/AlO\scriptsize x\normalsize \:).(b) Derivative of $\Delta$R\scriptsize f\normalsize \: vs. the field amplitude for a Pt/Co/Pt sample (inset Pt/Co/AlO\scriptsize x\normalsize \:).}
\label{1}
\end{figure}

In the presence of current induced displacements, the nonlinearities observed in the $\Delta$R\scriptsize f\normalsize\:versus field amplitude curve should be asymmetric. Moreover the asymmetry should depend on the current value. Such an asymmetry is observed (inset of figure 4) in the case of Pt/Co/AlO\scriptsize x\normalsize \: samples. In contrast to this behavior, a fully symmetric dependence that does not depend on the current amplitude is measured for the Pt/Co/Pt samples (figure 4). We conclude that in this case the spin torque induces DW displacements smaller than the resolution limit of this method. This limit value leads to (supplementary notes) $\beta$$\leq$0.02. Theoretical estimations\cite{zhang} based on a spin-flip frequency of 10$^{12}$ Hz yield a value $\beta$=0.01.

To clarify the difference of the spin-torque efficiency in the two samples, the symmetry breaking due to the presence of the AlO\scriptsize x\normalsize \: surface must be taken into account.
As a metallic film gets thinner, the conduction electron's behavior resembles more and more to that of a two-dimensional electron gas. When such a gas is trapped in an asymmetric potential well, the spin-orbit coupling is much stronger than in the case of a symmetric potential due to the Rashba interaction\cite{petersen-tb}. This effect was first evidenced in nonmagnetic materials where this interaction leads to a band splitting (0.15 eV for the surface states of Au (111)\cite{LaShell-Au}). In the case of ferromagnetic metals this effect was already proposed to contribute as an effective magnetic field\cite{tatara} for certain DW micromagnetic structures, but should not have any effect for Bloch walls in PMA materials. The simple 1D representation used in this case\cite{tatara} to model the DW accounts for the coherent rotation of the spins of the incoming electrons around the effective field, but excludes any de-coherence between electrons having different k-vector directions on the Fermi sphere (different directions of the Rashba effective field) as well as possible spatial inhomogeneities of this field (surface roughness). Since the spin-torque is caused by the cumulative action of all conduction electrons \cite{zhang}, the relevant parameter is not the spin-flip rate of a single electron but the relaxation rate of the out of equilibrium spin-density \cite{zhang}. In a more realistic 2D case, in the presence of the above mentioned strong decoherence effects, the relaxation rate of the out of equilibrium spin-density approaches the rate of spin precession around the Rashba effective field. The above value of the measured spin-orbit splitting (0.15 eV) will yield in this case an effective spin-flip rate of 30$\cdot$10$^{12}$ Hz, which is in excellent agreement with the order of magnitude of the measured non-adiabatic parameter, supporting this scenario.

In summary, a technique that allows the direct measurement of the torque from an electric current on a DW was developed. We have pointed out the importance of spin-flip interactions to spin torque by comparing its efficiency between two different systems. We show that the Pt/Co/AlO\scriptsize x\normalsize\; sample with the required symmetry properties to increase the spin-flip frequency (breaking of the inversion symmetry) shows an enhanced spin torque effect. A value of the order of 1 was measured for the $\beta$ parameter approaching the maximum value predicted by existing theories. This value can be explained by order of magnitude considerations on the Rashba effect observed on surface states of metals. Obtaining a high efficiency spin torque in a low coercivity material would make possible the development of nanoscale devices whose magnetization could be switched at low current densities. The order of magnitude of the current densities would be similar to the one observed for magnetic semiconductors\cite{yamanouchi-semi}, but, as the resistance is smaller, the supplied power will be lower.

\end{document}